\documentclass[aps,superscriptaddress,preprint,amsmath,amssymb]{revtex4}

\usepackage{graphicx,float}
\usepackage{dcolumn}
\usepackage{bm}

\usepackage{psfrag}

\begin{document}

\title{Mixtures of strongly interacting bosons in optical lattices}

\author{P. Buonsante}
 \affiliation{C.N.I.S.M. and Dipartimento di Fisica, Politecnico di Torino, C.so Duca degli Abruzzi 24, I-10129 Torino, Italy}%
\author{S.M. Giampaolo}
 \affiliation{Dipartimento di Matematica e Informatica, Universit\`a degli Studi di Salerno, Via Ponte don Melillo, 84084 Fisciano (SA), Italy}
 \affiliation{CNR-INFM Coherentia, Napoli, Italy, CNISM Unit\`a di Salerno,
and INFN Sezione di Napoli, Gruppo collegato di Salerno, Baronissi (SA), Italy}
\author{F. Illuminati}
 \affiliation{Dipartimento di Matematica e Informatica, Universit\`a degli Studi di Salerno, Via Ponte don Melillo, 84084 Fisciano (SA), Italy}
 \affiliation{CNR-INFM Coherentia, Napoli, Italy, CNISM Unit\`a di Salerno,
and INFN Sezione di Napoli, Gruppo collegato di Salerno, Baronissi (SA), Italy}
\affiliation{Institute for Scientific Interchange, Viale Settimio Severo 65, 10133 Torino, Italy}
\author{V. Penna}
 \affiliation{C.N.I.S.M. and Dipartimento di Fisica, Politecnico di Torino, C.so Duca degli Abruzzi 24, I-10129 Torino, Italy}%
\author{A. Vezzani}
 \affiliation{CNR-INFM and Dipartimento di Fisica, Universit\`a degli Studi di
 Parma, V.le G.P. Usberti n.7/A, I-43100 Parma, Italy}
\date{\today}

\begin{abstract}
We investigate the properties of strongly interacting heteronuclear boson-boson mixtures loaded in realistic optical lattices, with particular emphasis on the physics of interfaces. In particular, we numerically reproduce the recent experimental observation that the addition of a small fraction of $^{41}$K induces
a significant loss of coherence in  $^{87}$Rb, providing a simple explanation. We then investigate  the robustness against the inhomogeneity typical of realistic experimental realizations of the glassy {\it quantum emulsions} recently predicted to occur in strongly interacting boson-boson mixtures on ideal homogeneous lattices. 
\end{abstract}

\maketitle

\setlength{\parskip}{0pt}
\setlength{\parsep}{0pt}

Ultracold degenerate gases in optical lattices provide an
unprecedented toolbox for realizing experimentally what were once just
toy models sketching the key features of complicated condensed matter
systems. One prominent example is the Bose-Hubbard (BH) model,
originally introduced as a variant of the better known Hubbard model
\cite{Haldane_PLA_80_280} and later adopted for the description of
superfluid $^4$He trapped in porous media
\cite{Fisher_PRB_40_546}. Several years after the introduction of this
simple yet challenging toy model, Jaksch and co-workers  suggested
that it could be realized in terms of ultracold bosonic gases trapped
in optical lattices \cite{Jaksch_PRL_81_3108}, and were soon proved
right by a breakthrough experiment where the hallmark
superfluid-insulator quantum phase transition of the BH model was
observed \cite{Greiner_Nature_415_39}.

Recently, several experimental groups directed their efforts to the realization of more complex generalizations of the Hubbard model, involving  mixtures of particles obeying either the same or different statistics. Beyond their theoretical appeal, these systems are relevant to interesting  applications such as implementation of disorder \cite{Gavish_PRL_88_170406,Ospelkaus_PRL_96_180403}, association of dipolar molecules \cite{Damski_PRL_90_110401}, schemes for quantum computation \cite{Daley_PRA_69_022306} and mapping of spin arrays \cite{Spin}.

So far, most of the experimental efforts on optical lattice systems
have been directed to  boson-fermion mixtures
\cite{Gunter_PRL_96_180402,Ospelkaus_PRL_96_180403,Luhmann_CM_0711_2975},
while fermion-fermion and boson-boson (BB) mixtures have been somewhat
ignored. Very recently the Florence group performed an experiment
on a {\it harmonically trapped}
BB mixture of atomic $^{41}$K and $^{87}$Rb
with strong interspecies repulsion \cite{Catani_PRA_77_011603}.
Expectedly, the presence of a relevant K fraction modifies
the quantum phase transition occurring in Rb. More surprisingly, this
effect turns out to be sizeable even for a small overlap between
the two atomic species \cite{Catani_PRA_77_011603}. Strongly interacting
BB mixtures are also the subject of a recent theoretical investigation,
whose main observation is that strong interspecies repulsion can substitute for
disorder, driving a mixture loaded in a {\it homogeneous} 1-D lattice into
metastable {\it quantum emulsion} states exhibiting glassy features \cite{Roscilde_PRL_98_190402}.

In the present work we introduce a unified framework for the description
of lattice BB mixtures with strong interspecies interactions in realistic
conditions and different physical regimes encompassing and generalizing
the above-described findings \cite{Catani_PRA_77_011603,Roscilde_PRL_98_190402}.
In particular, we explain the apparently surprising observation that the coherence
properties of a bosonic system can be reduced significantly even in the presence
of a single interface with a second bosonic species \cite{Catani_PRA_77_011603}.
Furthermore, we establish the range of parameters for which the intuitively expected
opposite behavior of increased coherence is recovered.
Concerning {\it quantum emulsions}, we show that they are in principle
compatible with the inhomogeneity arising from confining potentials
typical of experimental realizations, albeit in a restricted range of
Hamiltonian parameters. Specifically, while in the homogeneous case a
sufficiently strong interspecies repulsion ensures the occurrence of quantum
emulsion states \cite{Roscilde_PRL_98_190402}, in the experimentally
relevant inhomogeneous case the difference of intraspecies repulsions
turns out to be a fundamental critical parameter.

The systems under concern provide a realization of the
two-flavor BH Hamiltonian 
\begin{eqnarray}
\label{E:sflv}
H &=& U_{1\,2} \sum_j n_{1,j}\, n_{2,j}+ \sum_{f,j}\left[\frac{U_f}{2} n_{f,j} (n_{f,j}-1)  \right. \nonumber     \\
&+& v_{f,j} n_{f,j} - \left. J_f \sum_{\ell \sim j}\left( a_{f,j}^\dag a_{f,\ell} +  a_{f,\ell}^\dag a_{f,j}\right)\right]
\end{eqnarray}
where the lattice boson operators $a_{f,j}^\dag$, $a_{f,j}$, and $n_{f,j} =
a_{f,j}^\dag \,a_{f,j}$, create, destroy and count atoms of type $f$ at 
site $j$. The parameters $U_f$,
$U_{1\,2}$ quantify the intra- and inter-species BB (repulsive)
interaction, $J_f$ is the hopping amplitude and $v_{f,j}=  m_f
\Omega_f(j-j_f^0)^2/2$ is the standard harmonic trapping potential felt by 
bosons of species $f$ at lattice site
$j$. By $m_f$, $\Omega_f$, and $j_f^0$, we denote, respectively, the mass,
the trapping frequency, and the minimum point of the harmonic potential $v_{f,j}$ of species $f$.

Since our aim is the study of strongly interacting mixtures, it is convenient and effective to adopt a mean-field approach based on the assumption that the ground state of the system is the product of on-site factors $|\Psi\rangle \!=\!\prod_j |\psi_j\rangle$, $|\psi_j\rangle = \sum_{n_1,n_2}\! c_{n_1,n_2}^{(j)}\!\left(a_{1,j}^\dag\right)^{n_1}\!\! \left(a_{2,j}^\dag\right)^{n_2}\! |\Omega \rangle$, where $|\Omega \rangle$ is the vacuum state, $a_{f,j}|\Omega \rangle=0$, and the coefficients $c_{n_1,n_2}$ are determined via energy minimization at fixed atomic populations $N_1$, $N_2$.
Owing to a much lower computational demand, this mean-field approach provides  qualitative results on systems that would be beyond the present capabilities of  more quantitative numerical methods, such as quantum Monte Carlo (QMC), density matrix renormalization group or time-evolving block-decimation algorithms.

Hamiltonians similar to that under examination have been considered previously \cite{Jaksch_PRL_81_3108,Damski_PRL_90_110401,Altman_NJP_5_113,Moore_PRA_67_041603,Chen_PRA_67_013606,Ziegler_PRA_68_053602,Spin,Isacsson_PRB_72_184507,Krutitsky,Kuklov_PRL_92_050402,Mathey_PRB_75_144510,Arguelles_PRA_75_053613,Mishra_PRA_76_013604}, possibly referring to two different internal states of the same bosonic species \cite{Spin,Chen_PRA_67_013606} to spin-1  \cite{Krutitsky} or dipolar bosons \cite{Damski_PRL_90_110401,Arguelles_PRA_75_053613}. Most of the previous work focuses on the phase diagram of homogeneous lattices, often adopting a mean-field approximation similar to ours \cite{Damski_PRL_90_110401,Altman_NJP_5_113,Chen_PRA_67_013606,Ziegler_PRA_68_053602,Isacsson_PRB_72_184507,Krutitsky}. However, our approach is characterized by some features that have not been considered in the literature, at least simultaneously. First of all, our mean-field is fully site-dependent, and does not reduce to an effective single-site theory. This allows us to describe phase-separated systems and to consider realistic harmonic trapping potentials. Furthermore, we fix the bosonic populations, $N_1$ and $N_2$, rather than the corresponding chemical potentials, $\mu_1$ and $\mu_2$. Again, this allows us to make  direct contact with experimentally relevant situations, and avoids the ``species depletion'' problem \cite{Chen_PRA_67_013606,Isacsson_PRB_72_184507},  
i.e. the vanishing of one atomic species from the minimum-energy state at fixed chemical potential.

The method is first applied to a situation reproducing
the experimental conditions in Ref.~\cite{Catani_PRA_77_011603}, where
 a bosonic mixture of Rb and K was loaded in an optical lattice.
The potentials trapping the two atomic species had the same 
$k_f=m_f \Omega_f$ but, since $m_1\neq m_2$, their minima were displaced in
the vertical direction:
$j_f^0 =2 g/ (\Omega_f^2 \lambda) $, where $g$ is
the gravitational constant and $\lambda/2$ is the optical lattice spacing
(henceforth the subscripts $1$ and $2$ will denote Rb and K, respectively).
An important consequence of the interplay between the ensuing asymmetry and
 the strong interspecies repulsion is the tendency towards full phase separation,
minimizing the number of interfaces between the two species.
In fact, in the Florence experiment the interspecies overlap  is estimated to be
limited to one lattice site in the vertical direction. Despite the occurrence of
a single phase interface, the effect of K on the coherence properties of Rb turns out
to be sizeable \cite{Catani_PRA_77_011603}.
More in detail, it has been observed that a modest quantity of K
(around 10\% of Rb) reduces the coherence of Rb significantly, moving
the superfluid-insulator transition point to smaller values of the
optical lattice depth $s$. The authors of Ref. \cite{Catani_PRA_77_011603}
also remark that a naive argument based on Ref.~\cite{Damski_PRL_90_110401}
results in a prediction opposite to the observed behavior: the presence of K increases the
local density of  Rb, which would cause an increase in the coherence of the latter.

This argument is indeed valid for most of the phase diagram of the BH
model describing an atomic cloud loaded in a {\it homogeneous} optical
lattice. However, clear exceptions are found in the proximity of
the Mott lobes, where an increase of the (local) density --or
chemical potential-- results in a sharp drop in the 
condensate fraction. Furthermore it should be emphasized that such phase
diagram describes a homogeneous system in the thermodynamic limit,
whereas here we are dealing with a finite and inhomogeneous
system. The site-dependent potential acts like a local
chemical potential for a system with fixed total population. As a
result, at sufficiently high ratios of interaction to kinetic energy,
configurations of the system can be found where superfluid and
Mott-insulating domains coexist
\cite{Batrouni_PRL_89_117203,Kashurnikov_PRA_66_031601}. The density
of the system assumes the so-called {\it wedding-cake} or {\it
ziggurat} profile, the plateaus corresponding to (quasi)
Mott-insulator domains. When the configuration is such that the
topmost plateau involves a fair number of sites, the density profile
responds to an increase in the total population according to a
predictable sequence. At first a dome-like essentially superfluid
structure appears on top of the highest plateau. Subsequently, the width and height
of this structure increase, leading to an increase in the system
coherence. When the tip of the dome gets too close to the next level
of the {\it ziggurat}, the dome flattens, its central part turning
gradually into a plateau. Correspondingly, there is a drop in the
overall coherence of the system \cite{Batrouni_PRL_89_117203}.

\begin{figure}[t!]
\begin{center}
\psfrag{V}{${\cal V}^{(1)}$}
\includegraphics[width=8.5cm]{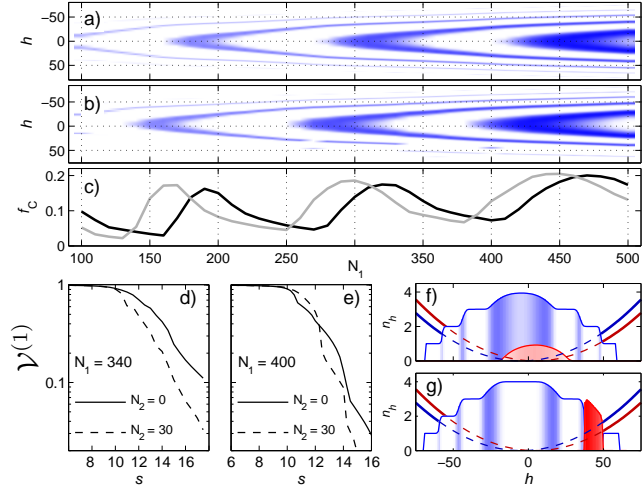}
\caption{\label{Fig1} (color online) a) $|\alpha_{1,h}|^2$ vs  $N_{1}$  ($s= 11$, $J_{1} \approx 0.022 \,U_1 $, $k_{1} \approx 7.9\cdot 10^{-4} U_1$); b) same as a), but in the presence  of K ($N_{2}=30$, $U_{2} \approx 0.65 \, U_1 $, $J_{2} \approx 0.21 \, U_1$, $U_{12} \approx 2.22 \,U_1$, $k_{2} = k_{1}$ c)  $f_{\rm C}$ of Rb corresponding to a) (black) and b) (gray); d) and e) visibility of Rb vs. lattice strength.  f) and g) Configurations of 340 Rb  (blue) and 30 K (red) atoms for $s=11$, along with the relevant trapping potentials (arbitrary units). The height of each bar represents the local population $n_h$, whereas the darkness of the shading is proportional to $|\alpha_{j,h}|^2$. In f) we set $U_{1\,2}=0$.}
\end{center}
\end{figure}

The above described single-species scenario is captured quite satisfactorily by the  Gutzwiller mean-field approximation \cite{Jaksch_PRL_81_3108,Pollet_PRA_69_043601,Zakrzewski_PRA_71_043601,NoteGW}.
We will now show that it bears a strict relation with the experimental observations reported in Ref.~\cite{Catani_PRA_77_011603} about the Rb-K BB mixture.
Fig. \ref{Fig1} shows results obtained from a double-species Gutzwiller mean-field approach where we have
adopted physical parameters  ---  $J_f$, $U_f$, $U_{1\,2}, \Omega_f$, atomic density at the trap center, population ratios --- in the experimentally determined range
\cite{Catani_PRA_77_011603,NotePars}.
For the sake of simplicity we have focused on a 1-D lattice as mean-field results are
essentially independent of the dimensionality \cite{NoteTilt}.
Panel a) shows the local superfluid parameter $|\alpha_{{1},h}|^2=|\langle\Psi|a_{{1},h}| \Psi\rangle|^2$ of  Rb alone as a function of the relevant population $N_{1}$ and lattice site label, $h$ (the darker the hue, the larger $|\alpha_{{1},h}|^2$). The drop in the superfluid parameter at the trap center signals the formation of new ziggurat levels from the flattening of coherent domes.
Panel b) shows the same quantity as in a) yet in the presence of 30 atoms of K ($|\alpha_{{2}, j}|^2$ is not shown). The main effect of the  addition of K is that the new structures of the (now asymmetric) ziggurat appear at smaller populations $N_{1}$. Panel c) shows an estimate of the coherence of Rb measured in terms of the relevant condensate fraction $f_{\rm C}^{1}$ \cite{Penrose_PR_104_576} for the data in panels a) (black) and  b) (gray). 
The presence of K is indeed equivalent to an increase in  Rb
population, but, given the oscillatory behavior of $f_{\rm C}^{1}$,
{\em this does not necessarily result in an increase of the overall
coherence of} Rb. A small fraction of $N_2$ can cause either an increase or a decrease of $f_{\rm C}^{1}$, depending on the value of $N_1$. The experimental
measure of coherence, i.e.  the so-called {\it visibility} $\cal V$ \cite{Gerbier_PRA_72_053606,Catani_PRA_77_011603}, exhibits similar oscillations as in c), albeit with a different envelope.  Panels d) and e) show the changes in ${\cal V}^{(1)}$ produced by $N_2 = 30$ K atoms, for two values of $N_1$. Note that d) considers the  same population ratio as estimated in the experiment
\cite{Catani_PRA_77_011603}, and  reproduces quite satisfactorily the observed loss of coherence. Guided by panel c), in e) we change $N_1$ from 340 to 400 to probe the opposite phenomenon. It turns out that the presence of K
 enhances ${\cal V}^{(1)}$ only at relatively low lattice depths, while at large $s$ the effect is again a loss of coherence, albeit less pronounced. This result agrees with experiments, where an increase of coherence was never observed \cite{NotePars}. 

\begin{figure}[t!]
\begin{center}
\includegraphics[width=8.5cm]
{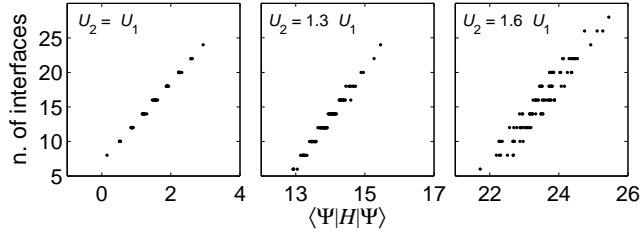}
\caption{\label{Fig2}  number of interfaces vs energy for $N_1=N_2=180$ particles on a 300-site 1D lattice. In all cases $J_1=J_2 = 0.1\, U_1$ and $U_{1\,2}=1.5\, U_1$.}
\end{center}
\end{figure}
We now turn our attention to another interesting feature of strongly interacting BB mixtures, i.e. the possible occurrence of low energy metastable states characterized by a large number of interfaces, recently discussed in the ideal case of homogeneous lattices, $\Omega_{f,j}=0$, \cite{Roscilde_PRL_98_190402}.
The authors of Ref.~\cite{Roscilde_PRL_98_190402} observe that the QMC simulations employed to determine the ground state of the total Hamiltonian $H$ fail to
equilibrate as soon as $U_{1\,2} \geq U_f$, and ascribe this behavior  to the presence of many low-energy metastable states (where {\it metastable} refers to the robustness of these configurations against the QMC minimization algorithm, which is equipped with nonlocal moves).
Being characterized by a large number of interfaces separating single-species {\it droplets}, these metastable states are dubbed {\it quantum emulsions}. The relevant energies are found to be linearly dependent on the number of interspecies interfaces. One interesting feature of these quantum emulsions is their {\it spontaneous randomness}, i.e. the fact that the droplets exhibit a disordered spatial arrangement despite the absence of any randomness in the Hamiltonian parameters.

Adopting a self-consistent dynamical search algorithm for the ground state of the {\em homogeneous}
system in its Gutzwiller form \cite{NoteMenotti}, we find that the BB mixture gets trapped into a quantum emulsion state whose energy depends on the number of interfaces, in complete analogy with the results
obtained in Ref.~\cite{Roscilde_PRL_98_190402}. This is evident from Fig.~\ref{Fig2} illustrating the situation on a homogeneous lattice for different values of the hopping to interaction ratios \cite{noteInterf}.  However, the homogeneous lattice of  Ref.~\cite{Roscilde_PRL_98_190402} is a strongly idealized situation, in which the only requirement for the occurrence of quantum emulsions is that $U_{1\,2}$ be sufficiently larger than $U_1$ and $U_2$ \cite{Roscilde_PRL_98_190402, Altman_NJP_5_113}.

\begin{figure}[t!]
\begin{center}
\includegraphics[width=8.5cm]{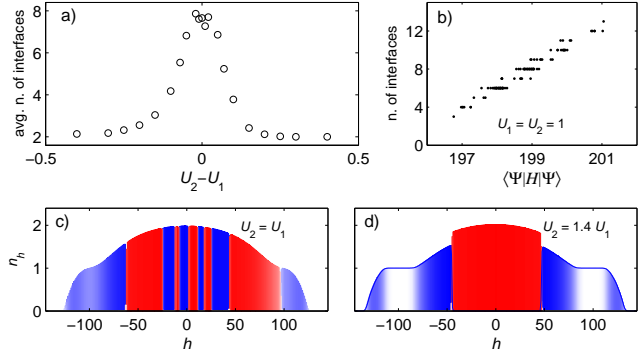}
\caption{\label{Fig3} (color online) Quantum emulsions in the presence of a harmonic confinement; a) average number of interfaces (over 100 realizations) vs intra-species interaction difference; b) same as Fig. \ref{Fig2}.a; c) and d)  local densities of bosons in two limiting situations. The color key is the same as  Fig.~\ref{Fig1} f) and g). In all cases $N_1\!=\!N_2\!=\!180$, $k_1=k_2=10^{-4}\,U_1$, other parameters as in Fig.~\ref{Fig2}. }
\end{center}
\end{figure}

Moving to the inhomogeneous case typical of actual experimental situations, we find that
$\Delta U=|U_1 -U_2|$ becomes a further critical parameter for the existence of quantum emulsions. This is clearly illustrated in Fig.~\ref{Fig3}. Panel a)  shows the average number of interfaces as a function of $\Delta U$, while the inset is the analogous of the leftmost panel in Fig.~\ref{Fig2}. Panels b) and c) show  typical configurations at small and large values of $\Delta U$, respectively. Note that the former is characterized by a significant number of randomly arranged single-species droplets. In this case $J_1=J_2=0.1 U_1$, but we obtain similar results also for $J_1\neq J_2$, provided that $\Delta U \approx 0$. Clearly the number of quantum emulsion states at a given energy will be smaller compared to the homogeneous case, owing to the reduced degree of symmetry of the system. Indeed, in this case, the energy of each droplet does depend on its position in the lattice, due to the local potential contribution.
Unlike the one dimensional case, in higher dimensions the interface energy of 
a droplet depends on its size. This fact, along with the larger lattice connectivity, is expected to hinder the occurrence of quantum emulsions.

In summary, we have investigated the properties of a strongly interacting bosonic mixture loaded into an optical lattice, going beyond the idealized situation of a homogeneous system. We considered the inhomogeneities arising from the presence of the harmonic trapping potential typical of standard experimental setups as well as from the differential gravitational sag originated by the difference in the masses of the two bosonic species.
We reproduced the apparently surprising results of the first experiment involving a BB mixture \cite{Catani_PRA_77_011603}, providing a simple explanation for the observed loss of coherence of $^{87}$Rb in the presence of a small fraction of $^{41}$K. Furthermore, our results predict that the opposite phenomenon, i.e. the increase of coherence predicted by the ``naive argument'' proposed in Ref.~ \cite{Catani_PRA_77_011603}, is limited to sufficiently shallow lattice depths.

We then investigated the effect of inhomogeneity on the {\it quantum emulsion} states formerly predicted on homogeneous lattices \cite{Roscilde_PRL_98_190402}. In particular, we found that, at variance with the homogeneous case, a large value of $U_{1\,2}$ is not sufficient for their occurrence, and a further
critical condition is a small value of $U_2-U_1$. This suggests that the use of Feshbach resonances could be a crucial ingredient for the experimental observation of quantum emulsions in heteronuclear mixtures. An intriguing alternative possibility for the realization of lattice
BB mixtures with directly built-in conditions $J_1 = J_2$ and $U_1 = U_2$
could be provided by a generalization of the models considered in Ref.~\cite{Arguelles_PRA_75_053613},
by considering dipolar bosons placed on two neighboring lattices with angular
relations such that the two sets of atoms interact via a strong interspecies
{\em repulsion} $U_{1\,2}$.

{\it Acknowledgments}. The authors acknowledge fruitful discussions with F. Minardi and J. Catani.


\begin{thebibliography}{99}
\bibitem{NoteGW}
Note that mean-field theory yields results  remarkably similar to Fig. 1 of Ref.
\cite{Batrouni_PRL_89_117203} by simply increasing the interaction strength
in order to compensate for the underestimation of the critical point
inherent in this approximation.
\bibitem{NoteTilt}
In order to obtain $|j_1^0-j_2^0|$ of the same order as in Ref. \cite{Catani_PRA_77_011603}
we assume that our 1-D lattice is tilted off the horizontal plane by $2^\circ$.
\bibitem{noteInterf}
The case where both species are insulating represents apparently an exception
  in this respect, due to the energetic degeneracy of the emulsion states.
  However, this can be resolved by resorting to suitable mappings
  \cite{Altman_NJP_5_113}. Also, we emphasize the high ideality of this
  situation, where $N_1+N_2$ must be commensurate with the lattice size.
\end{thebibliography}
\end{document}